\documentclass[12pt]{article}
\usepackage{graphicx}
\usepackage{graphicx}
\usepackage{amsmath}
\usepackage{amssymb}
\newcommand{\erggs}{\mbox{erg$\,$g$^{-1}$s$^{-1}$}}
\newcommand{\gccm}{\mbox{g$\,$cm$^{-3}$}}
\newcommand{\MS}{\mbox{$M_{\textstyle\odot}\,$}}
\newcommand{\NMS}[1]{\mbox{$#1\,M_{\textstyle\odot}$}}
\newcommand{\LS}{\mbox{$L_{\textstyle\odot}$}}
\newcommand{\NLS}[1]{\mbox{$#1\,L_{\textstyle\odot}$}}
\newcommand{\RS}{\mbox{$R_{\textstyle\odot}\,$}}
\newcommand{\NRS}[1]{\mbox{$#1\,R_{\textstyle\odot}$}}
\newcommand{\frat}[2]{\mbox{$\frac{
\raisebox{.2ex}{$\textstyle #1$}}{\raisebox{-.7ex}{$\textstyle #2$}}$}}
\newcommand{\xmn}[2]{\mbox{$#1\!\times\! 10^{#2}\,$}}
\newcommand{\nucmu}{\mbox{$m_{\mathrm u}$}}
\newcommand{\indis}[1]{{\mbox{\scriptsize #1}}}
\newcommand{\aceqva}{&\!\!\!\! =\!\!\!\! &}
\newcommand{\ctwo}{\mbox{$\!\! $}}
\newcommand{\cthree}{\mbox{$\!\!\! $}}
\newcommand{\cfour}{\mbox{$\!\!\!\! $}}

\begin{document}
\input epsf

\title{Similarity theory of stellar models and structure
       of very massive stars}

\author{D.K. Nadyozhin$^{1,2}$ \and T.L. Razinkova$^2$}

 \date{}
\maketitle
\begin{center}
 $^1${\em Max-Planck-Institut f\"ur Astrophysik, Garching, Germany}\\
 $^2${\em A.I. Alikhanov Institute for Theoretical and Experimental 
 Physics, Moscow, Russia}
\end{center}

\begin{abstract}
 The similarity theory of stellar structure is used to study the properties 
 of very massive stars when one can neglect all the sources of opacity 
 except for the Thomson scattering. The dimensionless internal structure 
 of such stars is practically independent of the energy generation law.
 It is shown that the mass-luminosity relation can be approximated by
 an analytical expression which is virtually universal regarding the chemical
 composition and energy generation law. A detailed comparison with the 
 Eddington standard model is given. The application of the obtained results
 to the observations of massive stars is briefly discussed.  
\end{abstract}
\section{Introduction}
 The similarity theory of stellar structure was a powerful
 tool to understand the properties of stellar models 
 during the pre-computer era of their study. 
 Suffice it to mention a theoretical
 explanation of the mass-luminosity and mass-radius relations
 (Biermann, 1931; Str\"omgren, 1936; Sedov, 1959). 
 For years the similarity theory remains to be useful for 
 the interpretation of various aspects of stellar structure
 (Schwarzschild, 1958; Chiu, 1968; Cox, Guili, 1968; Dibai, Kaplan, 1976;
 Kippenhahn, Weigert, 1990). 
 
 Here, first we describe the similarity theory of stellar structure
 as a boundary-value problem formulated by Imshennik and Nadyozhin (1968). 
 Then we discuss the structure of chemically homogeneous 
 stars of so large a mass that the opacity can be considered
 as being due to the Thomson scattering alone.
 Such an approximation is believed to be adequate for 
 still hypothetical Pop~III stars and for a number
 of observed luminous stars --- e.g., massive O-stars,
 Wolf-Rayet stars, and some specific stars like $\eta\,$Car, 
 and the Pistol star. A special consideration is given to the comparison
 with the Eddington standard model.
 
\section{Similarity theory of stellar models}
 Let us consider a spherical star in hydrostatic and thermal 
 equilibrium. If the pressure $P$ is contributed by perfect gas
 $P_{\mathrm g}$ and black body radiation $P_{\mathrm r}$
 and the rate of energy generation
 $\varepsilon$ and opacity $\kappa$ are the power functions
 of the temperature $T$ and density $\rho$ then the structure of 
 the star is described by the following set of differential 
 equations:
\begin{eqnarray}
 & &\hspace{-1.3cm} \frac{d P}{d r}= -\rho\,\frac{Gm}{r^2}\, ,
    \label{hydroP} \\
 & &\hspace{-1.3cm}\frac{d m}{d r} =4\pi r^2\rho\, ,
    \label{dmr}\\
 & &\hspace{-1.3cm}\frac{d L}{d r} = 4\pi r^2\rho\,\varepsilon\, ,
    \label{dlr}\\
 & &\hspace{-1.3cm}\frac{d T}{d r}=\left\{\parbox{10cm}{
 $-\frat{\rho\, T}{P}\frat{Gm}{r^2}\,\nabla_A\,$, \quad
    $\nabla_r \geqslant\nabla_A\, ,\quad
     \nabla_A =\left(\frat{\partial\log T}{\partial\log P}\right)_S$ 
 ,\\
 $-\frat{\rho\, T}{P}\frat{Gm}{r^2}\,\nabla_r\,$, \quad
 $\nabla_r \leqslant\nabla_A\, ,\quad \nabla_r =
 \frat{3\kappa LP}{16\pi c\, aT^4\, Gm}$\, ,}\right. \label{dtr}
  \end{eqnarray}
 \begin{equation}
 P = P_{\mathrm g} + P_{\mathrm r} =
 \frac{k}{\nucmu}\frac{1}{\mu}\,\rho\, T + \frac{1}{3}\, a T^4\, ,
 \label{Ptot}
 \end{equation}
 \begin{equation}
 \beta =\frac{P_{\mathrm g}}{P}\, ,\qquad  \beta^{-1}= 1 + \mu\,\frac{a\nucmu}{3k}\,
 \frac{T^3}{\rho},\qquad \nabla_A =\frac{2(4-3\beta)}{32-24\beta -3\beta^2}
 \, ,\label{betadef}
 \end{equation}
  \begin{equation}
   \kappa =\kappa_0\,\rho^\alpha\, T^{-\nu}\, , \quad
     \varepsilon = \varepsilon_0\,\rho^\delta\, T^\eta\, , \label{KE}
  \end{equation}
  where $\beta$ is the ratio of the gas pressure to the total one; 
  $G$, $k$, and $a$  are the gravitational, Boltzmann, and radiation 
  density constants, respectively; \nucmu\ is the atomic mass unit
  and $c$ is the speed of light. The mean molecular mass $\mu$ and 
  the coefficients $\kappa_0$ and $\varepsilon_0$ depend on composition.
  
  The above equations should meet the boundary conditions:
 \begin{eqnarray}
 & \mbox{Center}\,\quad r=0\, : &\qquad m=0\, ,\;\; L=0\, ,\label{boundc}\\[2mm]
 & \mbox{Surface}\quad r=R : &\qquad P=0\, ,\;\; \rho=0\, ,\;\; m=M\, .
 \label{bounds}  
 \end{eqnarray}
 The conditions (\ref{boundc}) tell that  there is neither point mass nor 
 point source of energy in the center of the star. The conditions (\ref{bounds})
 at stellar surface state that both the pressure and density vanish there
 and the mass $m$ must be equal to the total mass $M$ specified for the star.
 The stellar radius $R$ is to be obtained as a result of the solution of 
 Eqs.$\,$(\ref{hydroP})--(\ref{KE}), i.e. $R$ is an eigenvalue of the problem.
 Simultaneously, the solution gives the luminosity of the star
 $L_0\equiv L(R)$.
 
 Assuming that the chemical composition and, consequently, $\mu$, 
 $\kappa_0$, and $\varepsilon_0$ are constant throughout the star,
 one can reduce the above equations to a dimensionless form by measuring
 the physical quantities in the following set of units 
 (Schwarzschild, 1958):
\begin{eqnarray}
 & & r \;\rightarrow\; R\, ,\qquad  m\;\rightarrow\; M\, ,
  \qquad  L \;\rightarrow\; L_0\, ,\nonumber\\[-3mm]
  & & \label{units}\\
  & & T \,\rightarrow\, \mu\frac{\nucmu}{k}
       \frac{GM}{R}\equiv T_0\, ,\quad
  P \,\rightarrow\,\frac{GM^2}{4\pi R^4}\equiv P_0\, ,\quad \rho\,
  \rightarrow\,\frac{M}{4\pi R^3}\equiv\rho_0\, .\nonumber
\end{eqnarray}
 
\noindent Introducing the following dimensionless variables
\begin{equation}\label{dimlessvar}
 \cthree x =r/R,\quad\! q=m/M,\quad\! l=L/L_0,\quad\! 
 p=P/P_0, \quad\!\sigma =\rho/\rho_0,\quad\! t=T/T_0\, ,
\end{equation}
 one can rewrite Eqs. (\ref{hydroP})--(\ref{bounds}) 
 in the form (Imshennik, Nadyozhin, 1968):
\begin{eqnarray}
 & &\frac{dp}{dx}= -\frac{\sigma q}{x^2}\, ,\label{hydroPd}\\
 & &\frac{dq}{dx}= x^2\sigma\, ,\label{dmrd}\\
 & &\frac{dl}{dx}=C_1\, x^2\sigma^{1+\delta}t^{\eta}\, ,\label{dlrd}\\
 & &\frac{dt}{dx}=\left\{
 \parbox{10cm}{$-\frat{q\sigma t}{x^2p}\,\nabla_A\,$, \quad
 $\nabla_r \geqslant\nabla_A\,$ ,\quad 
 $\nabla_A(\beta)=\frat{2(4-3\beta)}{32-24\beta-3\beta^2}\,$,\\[3mm]
 $-\frat{q\sigma t}{x^2p}\,\nabla_r\,$, \quad $\nabla_r\leqslant\nabla_A\,$,
   \quad $\nabla_r = C_2\,\frat{p\, l\,\sigma^\alpha}{q\, t^{4+\nu}}\,$,
   }\right. \label{dtrd}\\
 & & p=\sigma t +B\, t^4\, ,
     \qquad \beta^{-1}=1+B\,\frac{t^3}{\sigma}\, ,\label{Pbetd}
  \end{eqnarray}
  where all the parameters are gathered in the three dimensionless 
  constants:
\begin{eqnarray}
C_1 & = & \frac{1}{(4\pi)^\delta}\left(\frac{G\nucmu}{k}\right)^\eta
 \mu^\eta M^{1+\delta+\eta}\;\frac{\varepsilon_0}{L_0R^{3\delta+\eta}}
 \, ,\label{C1}\\
C_2 & = & \frac{3\,(4\pi)^{-\alpha}}{64\pi^2\, ac}
  \left(\frac{k}{G\nucmu}\right)^{4+\nu}\mu^{-4-\nu}M^{\alpha-3-\nu}
 \frac{\kappa_0 L_0}{R^{3\alpha-\nu}}\, ,\label{C2}\\
 B & = & \frac{4\pi a}{3G}\left(\frac{G\nucmu}{k}\right)^4 
\left(\mu^2 M\right)^2\, =\, 0.78096\left(\mu^2M/\MS\right)^2 .\label{B}
\end{eqnarray}
 
 \noindent The dimensionless boundary conditions become:
\begin{eqnarray}
 & & \mbox{Center $\; x=0$}:\quad q=0\, ,\;\; l=0\, ,
   \label{boundcd}\\
 & &\mbox{Surface $x=1$}:\quad p=0\, ,\;\; t=0\, ,
     \;\; l=1\, ,\;\; q=1 \, .\label{boundsd}
 \end{eqnarray}
 
 Thus, we have now six boundary conditions for four first order 
 differential equations (\ref{hydroPd})--(\ref{dtrd}). 
 This means that for every given
 $B$ (or $\mu^2 M$) the constants $C_1$ and $C_2$ can have only quite 
 certain values (eigenvalues) in order that the solution of {\em four\/}
 differential equations would satisfy to {\em six\/} boundary conditions.
 The solution itself as well as $C_1$ and $C_2$ depend only on $\mu^2 M$
 and the exponents $\alpha$, $\nu$, $\delta$, $\eta$:
 $C_{1,2}=C_{1,2}(\mu^2M,\alpha ,\nu , \delta, \eta)$.
 
 Equations (\ref{hydroPd})--(\ref{Pbetd}) can be solved by standard 
 methods. For the calculations described in the next section, 
 we have used the following approach. Having chosen some trial values
 of $C_1$ and $C_2$ and integrating Eqs. (\ref{hydroPd})--(\ref{Pbetd})
 numerically from the surface [$x=1$, boundary conditions (\ref{boundsd})]
 inward to some value $x=x_f$ $(0<x_f<1)$, one obtains a set of four
 quantities $\left\{p(x_f),q(x_f),l(x_f),t(x_f)\right\}$ as functions of
 $C_1$ and $C_2$.
 Then integrating these equations again from the center 
 [$x=0$, boundary conditions (\ref{boundcd})] 
 with the same $C_1$ and $C_2$ and trial values $p(0)=p_c$ and $t(0)=t_c$ 
 outward to $x_f$ one obtains a similar set of quantities that depends now on 
 $C_1$, $C_2$, $p_c$, and $t_c$. Since $p(x)$, $q(x)$, $l(x)$, and $t(x)$ 
 must be continuous throughout the star, both the sets have to coincide 
 at $x=x_f$. 
 One can organize an iterative process for four unknown parameters
 $C_1$, $C_2$, $p_c$, and $t_c$ to find such their values that would
 ensure the desired continuity of four variables 
 $p$, $q$, $l$, and $t$ at $x=x_f$. Of course, the result does not 
 depend on $x_f$ but the convergence domain of the iterations does. 
 
 As soon as $C_1$ and $C_2$ are found, one can easily derive 
 the luminosity $L_0$ and radius $R$ from Eqs.$\,$ (\ref{C1}) 
 and (\ref{C2}) which represent the mass-radius and mass-luminosity 
 relation, respectively.

\section{Structure of very massive stars}
In this section, the above similarity theory is employed 
to describe the structure of very massive stars.
In this case the opacity is dominated by the Thomson scattering
and one can assume $\alpha=0$, $\nu=0$, and 
$\kappa =\kappa_0=0.2(1+\mathrm{X})\;$cm$^2$/g 
(X is the mass fraction of hydrogen).

 The mass-luminosity relation [Eq. (\ref{C2})] can be converted
 into the form
 \begin{equation}
 L_0\, =\, \frac{64\pi^2\, ac}{3\,\kappa_0}
 \left(\frac{G\nucmu}{k}\right)^4\mu^4\, M^3\, C_2(\mu^2M,\delta,\eta)
 \, .\label{MLR}
  \end{equation}
 All the models with physically reasonable values of the energy generation
 temperature exponent (i.e., $\eta\geqslant 4$) considered below have convective 
 cores and radiative outer envelopes. Using Eqs.$\,$(\ref{hydroPd}) and 
 (\ref{dtrd}) and boundary conditions (\ref{boundsd}), one can find
 that $C_2$ is simply connected with 
 $\beta_{\mathrm s}(\mu^2M,\delta,\eta)$ --- the value 
 of the parameter $\beta$ at stellar surface:
 \begin{equation}
 1-\beta_{\mathrm s}\, =\, 4\, B\, C_2\, .\label{BsL}
 \end{equation} 
 Now the mass-luminosity relation can be reduced to
\begin{equation}
 L_0\, =\, L_{\mathrm{Ed}}(1-\beta_{\mathrm s})\, ,\label{Lbetas}
\end{equation}
  where the Eddington critical luminosity $L_{\mathrm{Ed}}$ is given by
  \begin{equation}
 L_{\mathrm{Ed}}\,\equiv\,\frac{4\pi c GM}{\kappa_0}\, =\,
 \frac{\xmn{6.483}{4}}{1+\mbox{X}}\frac{M}{\MS}\,\LS\, .\label{Ledd}
 \end{equation}
 
  Using the Runge--Kutta method with automatic control of accuracy
  for solving Eqs.$\,$(\ref{hydroPd})--(\ref{Pbetd}) from stellar surface
  down to $x_f$ and from center up to $x_f$, we have calculated a large
  number of models for a wide interval of the parameter $\mu^2M$
  $(0\leqslant\mu^2M\leqslant\NMS{4,000})$. 
  Typically, the values $x_f\approx 0.1-0.3$ are the best for 
  the iterations to converge. 
  The convergence domain, however, turned out to be rather narrow 
  at least for the Newton--Raphson iteration scheme used
  in our calculations.  So, when calculating a sequence of such models
  one has to change $\mu^2M$ at most by a few percents in order
  to ensure the convergence. 
  
  Table \ref{struct} presents
  the most important properties of several selected models.
  Note that for the sake of a better perception we use hereafter 
  for dimensionless quantities the same designations as 
  for dimensional ones.
  The first row of Table~\ref{struct} gives $\mu^2M$ with M measured
  in \MS. The next three rows are the $C_1$-values for three modes
  of energy generation: the CNO-cycle, $3\alpha$-reaction, 
  and pp-chain. The fifth row gives the $C_2$-values.
  The next six rows relate to the central dimensionless values of 
  density $\rho_{\mathrm c}$, pressure $P_{\mathrm c}$,
  temperature $T_{\mathrm c}$, 
  radiative temperature gradient $\nabla_{r\mathrm{c}}$,
  gravitational potential $\varphi_{\mathrm{c}}$ 
  (in the units $GM/R$), and ratio $\beta_{\mathrm{c}}$ of the gas 
  pressure to the total one. 
  The next two rows contain $\beta_{\mathrm{conv}}$ and $\beta_{\mathrm{s}}$
  at the convective core boundary and at stellar surface, respectively.
  Finally, the last six rows present the dimensionless values of radius, 
  mass, luminosity, density, pressure, and temperature at the 
  convective core boundary.
 \begin{table}
  \centering
  \caption{Structural properties of selected models}\label{struct}
  \vspace*{2mm}
 \begin{tabular}{l|lllllll}\hline\hline
 $\mu^2M$\cfour              & \multicolumn{1}{c}{0}\ctwo & \multicolumn{1}{c}{10} & \multicolumn{1}{c}{30} & \multicolumn{1}{c}{100} & \multicolumn{1}{c}{300} & \multicolumn{1}{c}{1000}\cthree & \multicolumn{1}{c}{4000}\\ \hline 
 \ctwo$\log_{10}C_1\, ^\star$\ctwo   & 1.6347\ctwo & 3.4318 & 5.8102 & 8.8083 & 11.768 & 15.281\cthree & 19.622\\ 
 \ctwo$\log_{10}C_1\, ^\dagger$\ctwo  & 2.1212\ctwo & 5.8346 & 10.615 & 16.602 & 22.509 & 29.523\cthree & 38.199\\ 
 \ctwo$\log_{10}C_1\, ^\spadesuit$\ctwo  & -0.4145\ctwo & 0.0220 & 0.5988 & 1.2961 & 1.9789 & 2.8051\ctwo & 3.8511\\ 
 \ctwo$\log_{10}C_2$\ctwo        & -3.2981\ctwo & -3.5181 & -3.9842 & -4.7533 & -5.5839 & -6.5626\cthree & -7.7304\\
 \ctwo$\rho_{\mathrm{c}}$\ctwo   & 59.34\ctwo & 60.20 & 65.05 & 79.97 & 98.95 
 & 119.4\cthree  & 137.5\\
 \ctwo$P_{\mathrm{c}}$\ctwo      & 45.55\ctwo & 43.95 & 46.43 & 58.41 & 75.32 
 & 94.78\cthree & 112.8\\
 \ctwo$T_{\mathrm{c}}$\ctwo & 0.7677\ctwo & 0.5816 & 0.4097 & 0.2630 & 0.1699 
 & 0.1014\cthree & 0.0539\\
\ctwo$\nabla_{r\mathrm{c}}\, ^\star$\ctwo & 2.459\ctwo & 3.249  & 4.525  & 5.806 &    6.588 & 6.992\cthree & 7.266\\
 \ctwo$\varphi_{\mathrm{c}}$\ctwo& -3.428\ctwo & -3.389 & -3.421 & -3.602 & -3.821 
 & -4.034\cthree & -4.204\\
 \ctwo$\beta_{\mathrm{c}}$\ctwo  & \multicolumn{1}{c}{1}& 0.7967 & 0.5739 & 0.3601 & 0.2232 & 0.1278\cthree & 0.0657\\
 \ctwo$\beta_{\mathrm{conv}}$\ctwo& \multicolumn{1}{c}{1} & 0.8671 & 0.6689 
 & 0.4294 &   0.2600 & 0.1429\cthree & 0.0704\\
 \ctwo$\beta_{\mathrm{s}}$\ctwo  & \multicolumn{1}{c}{1} & 0.9053 & 0.7085 & 0.4489 & 0.2673 & 0.1451\cthree & 0.0709\\
 \ctwo$r_{\mathrm{conv}}$\ctwo & 0.2832\ctwo & 0.3903 & 0.5107 & 0.6264 & 0.7077 
 & 0.7771\cthree & 0.8380\\
 \ctwo$m_{\mathrm{conv}}$\ctwo & 0.3120\ctwo & 0.5691 & 0.8063 & 0.9412 & 0.9823 
 & 0.99531\cthree & 0.99895\\
 \ctwo${L_{\mathrm{conv}}}^\spadesuit$ \ctwo & 0.7810\ctwo & 0.9799 & 0.9991 & 
 1.0000 & 1.0000  & 1.0000\ctwo & 1.0000\\
 \ctwo$\rho_{\mathrm{conv}}$\ctwo & 31.39\ctwo & 16.02 & 5.985 & 1.661 & 0.5056 
 & 0.1429\cthree & 0.0379\\
 \ctwo$P_{\mathrm{conv}}$\ctwo & 15.76\ctwo & 5.830 & 1.446 & 0.2540 & 0.05324 
 & 0.01049\cthree & 0.00184\\
 \ctwo$T_{\mathrm{conv}}$\ctwo & 0.5022\ctwo & 0.3157 & 0.1616 & 0.06564 & 0.02736 
 & 0.01036\cthree & 0.00342\\
 \hline\hline
 \multicolumn{8}{l}{\footnotesize $^\star\,$CNO-cycle $(\delta=1,\,\eta=16)$,\;
   $^\dagger\,$$3\alpha$-reaction $(\delta=2,\,\eta=30)$,\; 
  $^\spadesuit\,$pp-chain $(\delta=1,\,\eta=4)$}
\end{tabular}
 \end{table}
 
 The calculated sequences of one-parametric (depending on $\mu^2M$) 
 dimensionless models have so large convective cores that the major 
 part of thermonuclear energy turns out to be released within
 the core. Therefore, $L(r)$ becomes almost equal to 
 the luminosity $L_0$ everywhere outside the convective core.
 For such a strong dependence of the energy generation on temperature 
 as in the CNO-cycle $(\eta=16)$ and in $3\alpha$-reaction $(\eta=30)$, 
 one can with a high accuracy assume $l(x)=1$ in Eq. (\ref{dtrd}).
 Consequently, Eq. (\ref{dlrd}) for $l(x)$ proves to be
 disentangled from other equations (\ref{hydroPd}), (\ref{dmrd}),
 (\ref{dtrd}), and (\ref{Pbetd}). Now one can calculate the overall 
 stellar structure by solving a truncated eigenvalue problem that pays 
 no attention to the law of energy generation and considers the constant 
 $C_2$ as a single eigenvalue parameter involved.
 In particular, the resulting $C_2$-eigenvalue, and consequently $L_0$ 
 [Eq. (\ref{MLR})], as well as the dimensionless stellar structure
 do not depend on the exponents $\delta$ and $\eta$. Such a truncated
 model is actually the point-source model with  the Thomson opacity 
 and allowance for the radiation pressure that was first calculated 
 by Henrich (1943) within a limited range
 of stellar mass $(0\leqslant\mu^2M\leqslant 119)$. Our numerical
 results are in excellent agreement with his $\sim 1$\% accurate 
 calculations.
  \begin{figure}[ht]
 \centerline{\includegraphics[clip]{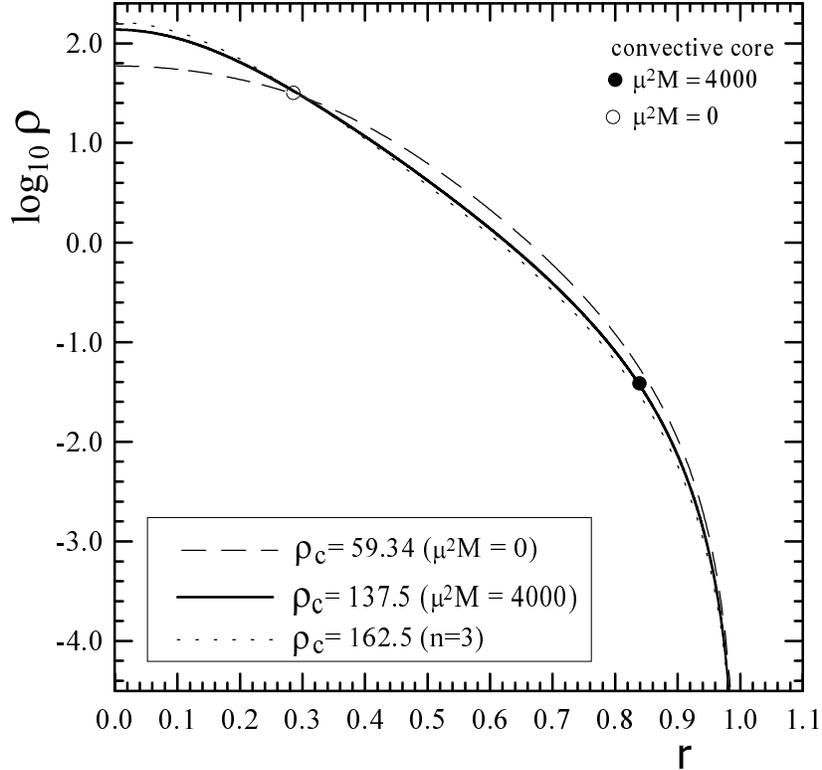}}
  \caption{Density versus radius in the units defined by 
  Eqs. (\protect\ref{units}) for $\mu^2M=0$ (dashed line),
  4000 (full line) and for a polytropic model ($n=3$, dotted line).
  The boundaries of the convective core are indicated for
  $\mu^2M=0\mbox{ and }4000$.}
  \label{dener}
\end{figure}
 
 When $C_2$ and the dimensionless functions $\sigma(x)$ and $t(x)$ 
 are already known from the solution of the truncated problem 
 one can find the constant $C_1$ and spatial distribution of 
 the dimensionless luminosity $l(x)$:
 \begin{equation}
 C_1=\left[\int_0^1\! x^2\,\sigma^{1+\delta}\, t^\eta\,{\mathrm d}x\right]^{-1},
 \qquad
 l(x)= C_1 \int_0^x\! x^2\,\sigma^{1+\delta}\, t^\eta\,{\mathrm d}x\, .
 \label{lc1}
 \end{equation}
 If the energy generation is an arbitrary 
 (but still strongly varying with temperature)
 function: $\varepsilon =\varepsilon(\rho,T)$, then instead of 
 $C_1$ one has to consider the stellar radius $R$ as an eigenvalue
 that can be found from the equation
  \begin{equation}
  L_0\, =\, M\int_0^1\! x^2\,\sigma(x)\;\varepsilon\!\left[
  \frac{M}{4\pi R^3}\sigma(x),\,\mu\frac{\nucmu}{k}
  \frac{GM}{R}t(x)\right]{\mathrm d}x\, ,\label{eigenR}
  \end{equation}
  where for given $M$ and composition the luminosity $L_0$ 
  is determined by the $C_2$-value obtained for the truncated model.
  
  The structural properties of the models presented in Table~\ref{struct}
  correspond to three modes of energy generation: the CNO-cycle, 
  $3\alpha$-reaction, and pp-chain. For the CNO-cycle and 
  $3\alpha$-reaction, the dimensionless luminosity 
  $L_{\mathrm{conv}}$ at the convective core boundary is equal to 
  1.0000 for all the values of $\mu^2M$. The same is true for the
  pp-chain when $\mu^2M\gtrsim 30$. All other parameters in 
  Tables~\ref{struct} and \ref{integr}, except for
  $C_1$ and $\nabla_{r\mathrm{c}}$, prove to be the same for the CNO-cycle
  and $3\alpha$-reaction, and at $\mu^2M\gtrsim 30$ for the pp-chain
  as well. Practically, one can consider only 
  the interval $0\leqslant\mu^2M\lesssim 10$ as exhibiting 
  slight differences for the pp-chain shown in 
  Figures \ref{rhopcen},$\,$\ref{convec},$\,$\ref{c1mod} below.
  This concerns also the mass-luminosity relation.
  More specifically, for $\eta=4$: $\log_{10} C_2 =-3.516$ and $-3.285$
  at $\mu^2 M =10$ and $0$, respectively.
\begin{figure}[ht]
  \centerline{\includegraphics[clip]{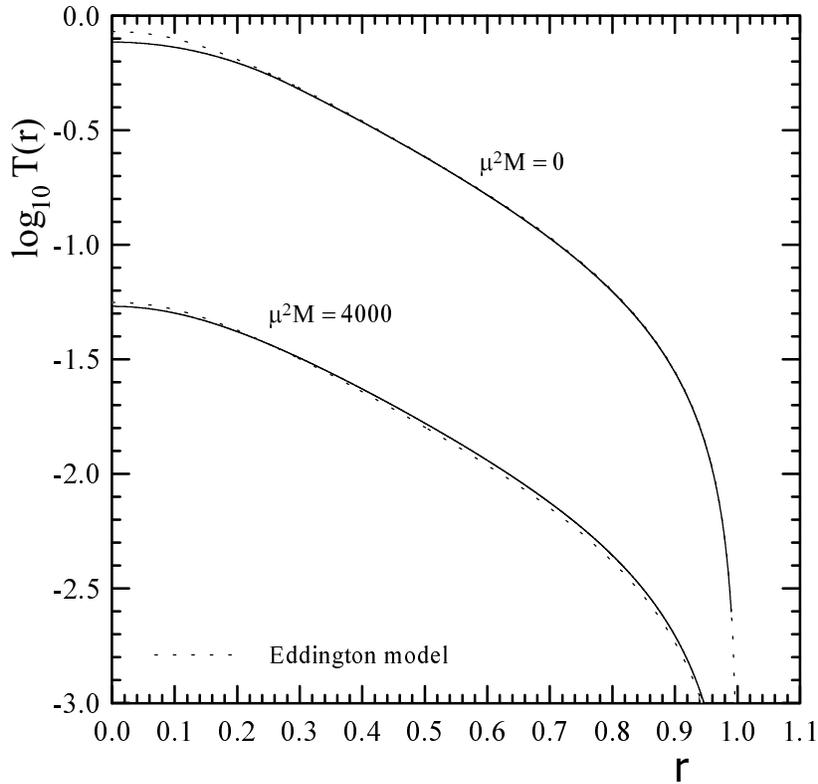}}
  \caption{Temperature versus radius in the units defined by 
  Eqs. (\protect\ref{units}) for $\mu^2M=0$ and 4000. 
  The Eddington model is shown by the dotted lines.}
  \label{temper}
\end{figure} 

  The constant $C_1$ and central value
  of the logarithmic radiative temperature gradient 
  $\nabla_{r\mathrm{c}}=C_1 C_2\, p_{\mathrm{c}\,}
  \sigma_{\mathrm{c}}^{\delta\,}t_{\mathrm c}^{\eta-4}$
  strongly depend on the energy generation law. The $\nabla_{r\mathrm{c}}$
  values for the CNO-cycle in Table~\ref{struct} demonstrate that even 
  for low masses $(\mu^2 M\lesssim 1)$ when one can neglect the radiation
  pressure, there is still a good excess of $\nabla_{r\mathrm{c}}$ over
  $\nabla_A$ $(0.25\leqslant\nabla_A\leqslant 0.4)$ 
  to drive the adiabatic convection 
  along stellar core. For the $3\alpha$-reaction, the excess is much larger:
  $\nabla_{r\mathrm{c}}=11.05$ for $\mu^2 M =0$. However for the pp-chain 
  we have $\nabla_{r\mathrm{c}}=0.594$ --- the value exceeding the adiabatic 
  gradient only by a factor of 1.5 $(\nabla_{A\mathrm{c}}= 0.4$ for 
  $\mu^2 M\approx 0)$. This happens because the exponent $\eta =4$ is not 
  very far from a critical value of $\eta\approx 2-3$ necessary to ensure 
  the existence of the convective core 
  (Cowling, 1934; Naur, Osterbrock, 1953).   

 Figures~\ref{dener} and \ref{temper} show the density and temperature
 versus radius for two extreme values of $\mu^2M$ in the case of the 
 CNO-cycle and $3\alpha$-reaction. With a high accuracy these two modes 
 of energy generation result in only one model for every given $\mu^2M$
 (see the discussion above). 

 It is instructive to compare the results with the Eddington standard
 model (Eddington, 1926) that at constant opacity 
 corresponds to a constant rate of energy 
 generation: $\varepsilon =\varepsilon_0$ $(\eta=0,\,\delta=0)$.
 In this case $C_1=1$ as it immediately follows from Eq. (\ref{lc1}).
 Then, it is easy to make sure that $\beta$ is to be constant inside
 such a model. As a result, the pressure turns out to be 
 connected with density along the radius by a power law: 
 \begin{equation}
  P\, =\, K\:\rho^{4/3},\qquad K\equiv\frac{k}{m_{\indis{u}}\mu}
  \left[\frac{3k}{am_{\indis{u}}\mu}\,\frac{1-\beta}{\beta^4}
  \right]^{1/3}=\mbox{ const}\, .
  \label{PEdd}
  \end{equation}
  \begin{figure}[ht]
  \centerline{\includegraphics[clip]{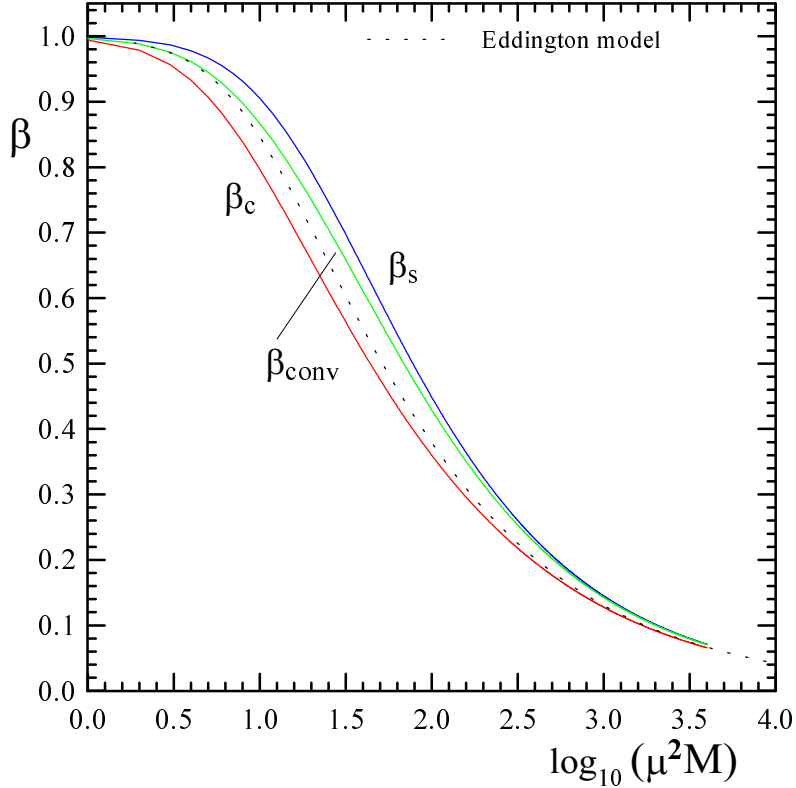}}
  \caption{The ratios of the gas pressure to the total one, $\beta$,
  at the center $\beta_{\mathrm{c}}$, the convective core boundary
  $\beta_{\mathrm{conv}}$, and the surface $\beta_{\mathrm{s}}$
  as functions of $\mu^2M$ ($M$ is in \MS). 
  A dotted curve corresponds to the
  Eddington standard model [Eq.~(\protect\ref{BetM})].}
  \label{betacc}
 \end{figure}
 Thus, the standard model is nothing else than a polytropic gas sphere
 of index $n=3$. The value of $\beta$ proves to be uniquely connected 
 with the mass $M$ by the following equation (Chandrasekhar, 1939):
  \vspace*{-1mm}
 \begin{equation}
 \frac{1-\beta}{\beta^4}\, =\, 0.01607\, G^3\left(\frac{m_{\indis{u}}}{k}
 \right)^4\left(\mu^2M\right)^2\, =\, \xmn{2.994}{-3}
 \left(\mu^2\frac{M}{\MS}\right)^2\, .
  \label{BetM}
  \end{equation}
 The constant $C_2$ as a function of $\mu^2M$ and the mass-luminosity 
 relation for the standard model are determined by Eqs.~(\ref{BsL}) and 
 (\ref{Lbetas}) with $\beta_{\mathrm{s}}$ being substituted for $\beta$ 
 from Eq.~(\ref{BetM}). It is easy to verify that the Eddington standard 
 model has no convective 
 core at all --- it is convectively stable for any $\mu^2M$.
 
 According to Fig.~\ref{betacc}, $\beta$ increases from the stellar 
 center up to the surface: 
 $\beta_{\mathrm{c}}<\beta_{\mathrm{conv}}<\beta_{\mathrm{s}}$, 
 the Eddington $\beta$ always remaining between $\beta_{\mathrm{c}}$
 and $\beta_{\mathrm{conv}}$. 
\begin{figure}[ht]
 \centerline{\includegraphics[clip]{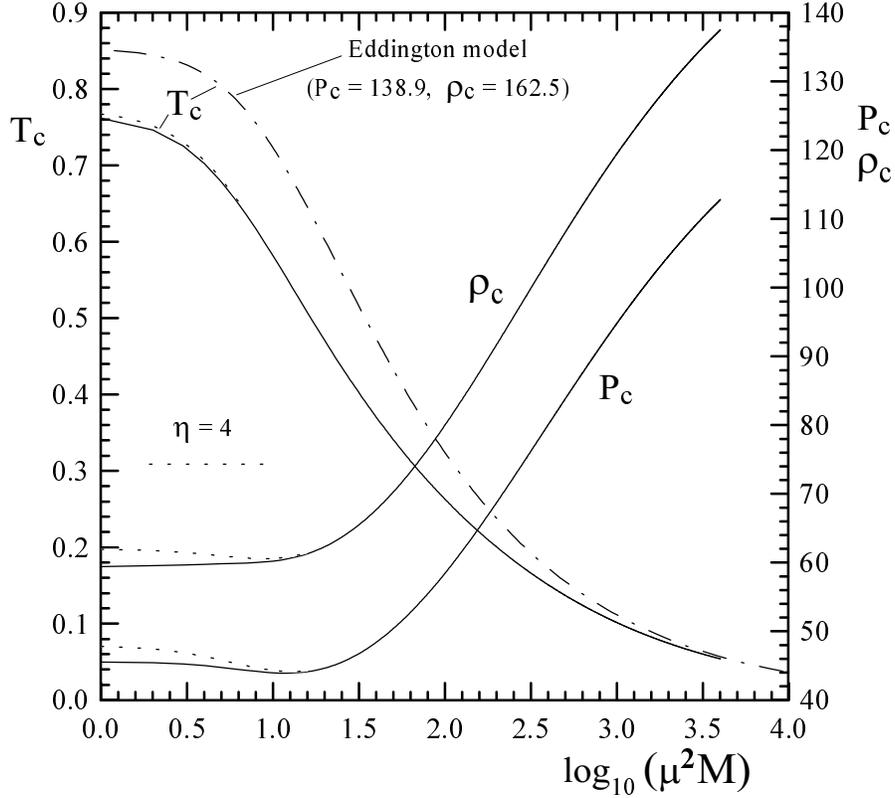}}
  \caption{The dimensionless central pressure $P_{\mathrm{c}}$,
  density $\rho_{\mathrm{c}}$ (right vertical axis), and 
  temperature $T_{\mathrm{c}}$ (left vertical axis)
  as functions of $\mu^2M$ ($M$ is in \MS).}
  \label{rhopcen}
\end{figure}
 
 Figure~\ref{rhopcen} shows the central pressure $P_{\mathrm{c}}$, 
 density $\rho_{\mathrm{c}}$ and temperature $T_{\mathrm{c}}$
 as functions of $\mu^2M$. At $\mu^2M\lesssim 10$, these quantities 
 for the pp-chain (dotted lines) only slightly differ from those for 
 the CNO-cycle and $3\alpha$-reaction (solid lines). The broken line
 represents $T_{\mathrm{c}}$ for the Eddington model for which 
 $P_{\mathrm{c}}$ and $\rho_{\mathrm{c}}$ are independent of $\mu^2M$:
 $P_{\mathrm{c}}=138.9$, $\rho_{\mathrm{c}}=162.5$ (polytrope $n=3\,$!).
\begin{figure}[ht]
  \centerline{\includegraphics[clip]{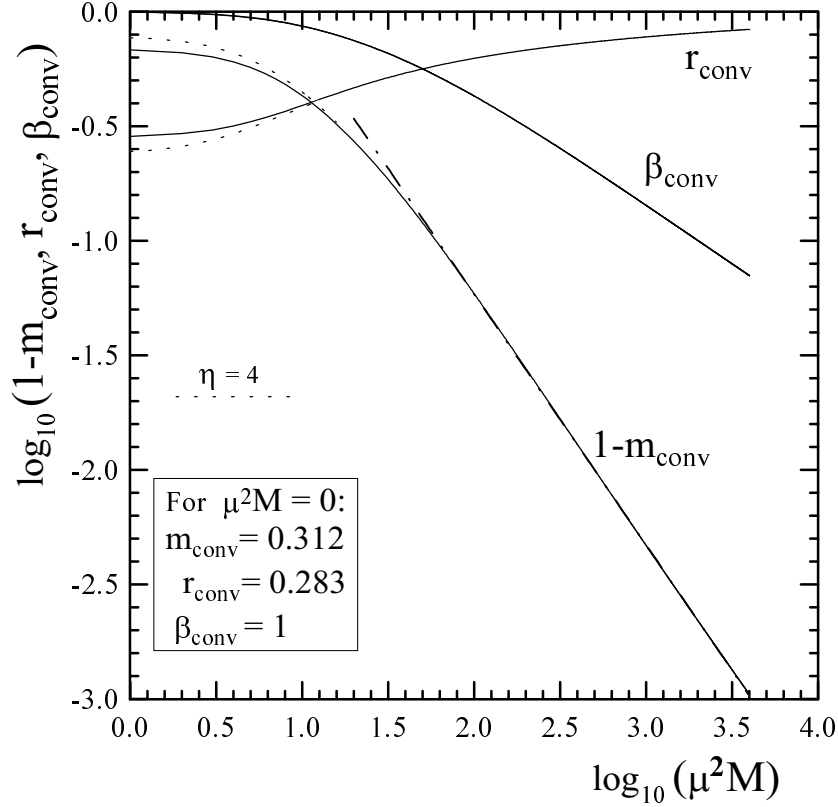}}
  \caption{The convective core boundary characteristics as functions
           of $\mu^2M$ for the CNO-cycle and $3\alpha$-reaction (solid lines)
           and pp-chain (dotted lines). The values for the CNO-cycle 
           and $3\alpha$-reaction at $\mu^2M=0$ are shown explicitly.}
  \label{convec}
\end{figure}

 Figure~\ref{convec} shows how the dimensionless radius of the convective 
 core $r_{\mathrm{conv}}$, the value of $\beta$ at the convective core 
 boundary $\beta_{\mathrm{conv}}$, and the mass above the convective core 
 $(1-m_{\mathrm{conv}})$ depend on $\mu^2M$. At $\mu^2M\gtrsim 40$,
 the convective core contains more than 85\% of stellar mass.
 One can approximate $m_{\mathrm{conv}}$
 (for all the three energy generation modes!) by the following 
 asymptotic relation (a straight broken line in Fig.~\ref{convec}):
 \begin{equation}
 m_{\mathrm{conv}}\, =\, 1 - 9.075\left(\mu^2M\right)^{-1.095}\, ,\quad
     \mu^2M\gtrsim 40\, .\label{mabov}
 \end{equation}
 \begin{figure}[ht]
  \centerline{\includegraphics[clip]{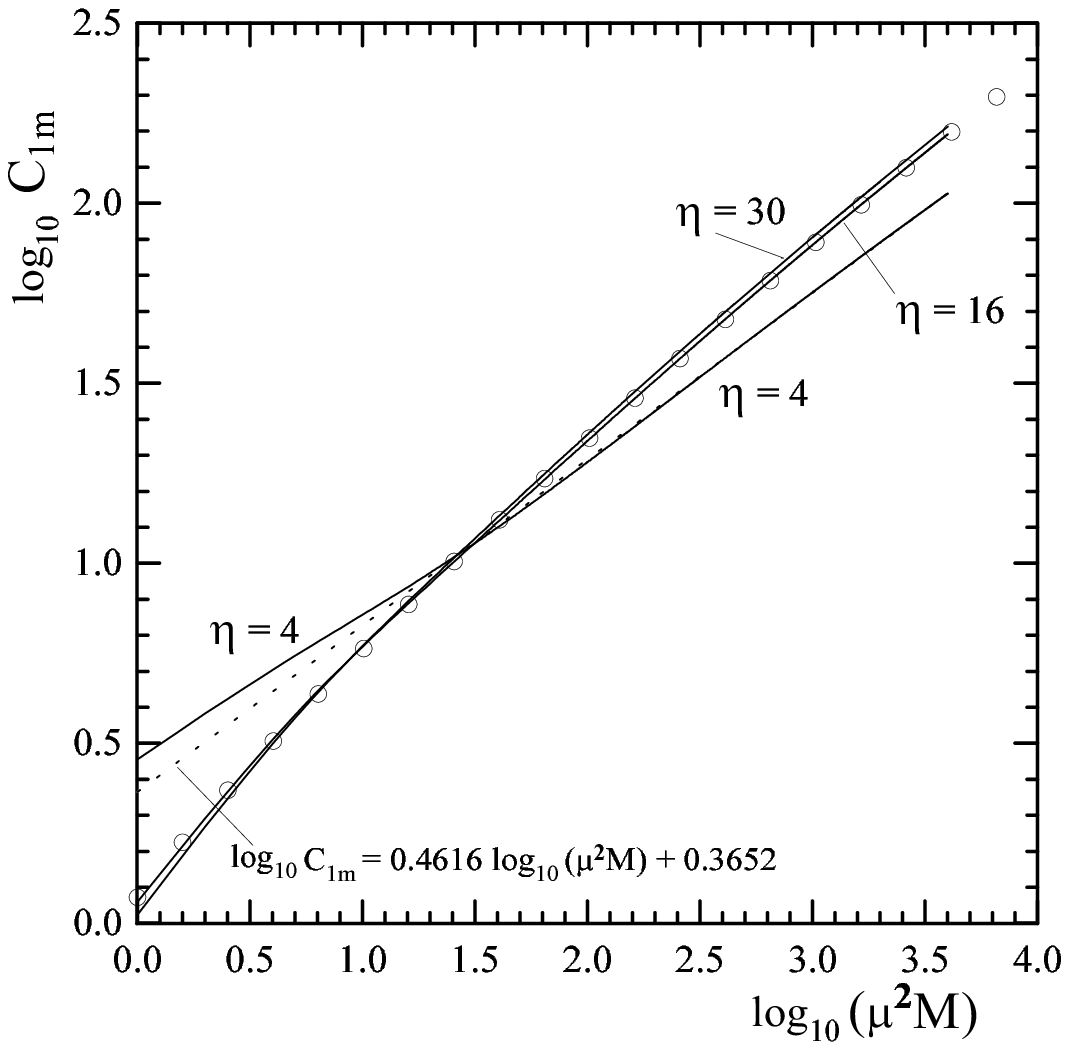}}
  \caption{The modified constant $C_{1\mathrm{m}}$ as a function of 
           $\mu^2M$ for the CNO-cycle, $3\alpha$-reaction and pp-chain.
           The open circles and dashed line show polynomial
           fits.}
  \label{c1mod}
\end{figure}

 Since the constant $C_1$ strongly depends on the energy generation
 mode (Table~\ref{struct}), it is useful to introduce a modified 
 constant $C_{1\mathrm{m}}$:
 \begin{equation}
 C_{1\mathrm m}(\mu^2M) \, \equiv\,
 (\mu^2M)^{\frac{\delta +\eta}{3\delta +\eta}}\,
 {\left[\left(1-\beta_{\mathrm s}\right)C_1\right]^{-\frac{1}{3\delta +\eta}}}
 \, ,\qquad (M\mbox{ in }\MS)
 \, .\label{C1mod}
 \end{equation}
 Now, using Eq. (\ref{Lbetas}) for $L_0$  we can obtain the following 
 expression for stellar radius as a function of $\mu^2M$ and composition:
 \begin{equation}
 \log R\, =\,\log C_{1\mathrm m}\, +\,\log D \, +\,\log S\, ,
 \quad (R\mbox{ in }\RS)\, ,\label{Rad}
 \end{equation}
 where 
 \begin{eqnarray}
\log D & = & \frac{1}{3\delta +\eta}\log [0.2\, (1+X)\, \xi]
  -\frac{2\delta +\eta}{3\delta +\eta}\log\mu\, ,\nonumber\\
\log S & = & \frac{\eta}{3\delta +\eta}\log\left(\frac{G\nucmu}{k}\right)
   -\frac{\delta +1}{3\delta +\eta}\log(4\pi)\nonumber\\
& & +\frac{1}{3\delta +\eta}\log\left(\frac{\varepsilon_{00}}{c\, G}
\right) +\frac{\delta +\eta}{3\delta +\eta}\log\MS\! -\log\RS\, ,
 \nonumber\\
 \xi & \equiv & \mathrm{X}^2\mbox{ \footnotesize (pp-chain), }\; 
 \mathrm{X}\,\mathrm{X}_{\mathrm{CNO}}\mbox{ \footnotesize (CNO-cycle), }\; 
 \mathrm{Y}^3\mbox{\footnotesize (3$\alpha$-reaction)}\nonumber\, .
\end{eqnarray}
 Here, the dependences of $\varepsilon_0$ on the mass fractions of 
 hydrogen $X$, the CNO-isotopes $\mathrm{X}_{\mathrm{CNO}}$, and helium $Y$
 are shown explicitly $(\varepsilon_0 = \varepsilon_{00}\;\xi)$.
 For large values of $\eta$, $C_{1\mathrm m}$ becomes independent
 of the energy generation mode --- 
 compare the CNO $(\eta=16)$ and $3\alpha$ $(\eta=30)$ curves in 
 Fig. \ref{c1mod}. Both the $C_{1\mathrm m}$
 can be approximated by a single polynomial shown by open circles
 in Fig. \ref{c1mod}:
 \begin{equation}
 y = 0.0720052 + 0.782547x - 0.120854x^2 + 0.0295923 x^3 - 0.0030574 x^4
 \, ,\label{LC1m}
 \end{equation}
 where $y\equiv\log_{10} C_{1\mathrm m}$ and 
 $x\equiv\log_{10}\left(\mu^2M\right)$. A fit for the pp-chain 
 $(\eta=4)$ is shown in Fig. \ref{c1mod} by a dashed line 
 that gives a good accuracy for $\mu^2M\gtrsim 10$.  
\begin{figure}[ht]
\centerline{\includegraphics[clip]{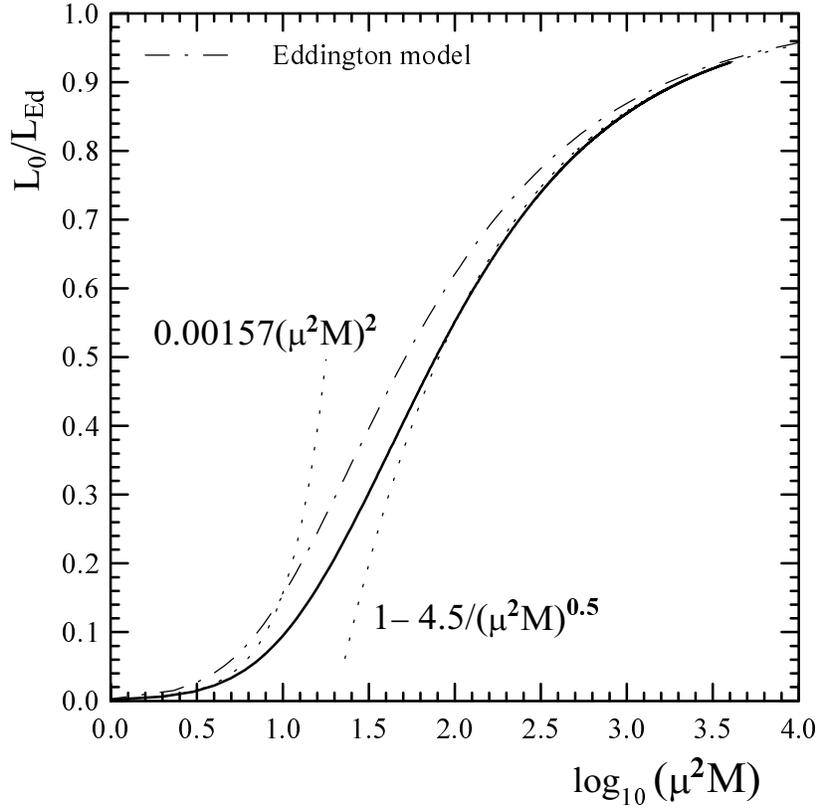}}
  \caption{The normalized mass-luminosity relation [Eq. (\protect\ref{Lbetas})].
  Dotted curves indicate asymptotics for small and large $\mu^2M$-values.
  The Eddington model is shown by a broken line.}
  \label{ledd}
\end{figure}

\begin{figure}[ht]
   \centerline{\includegraphics[clip]{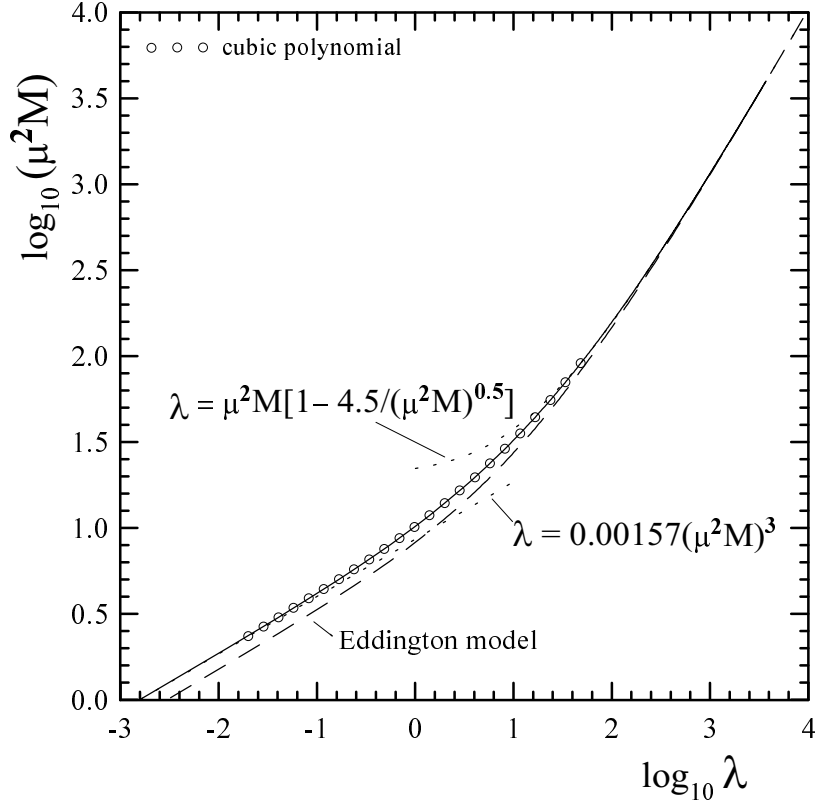}}
  \caption{The mass-luminosity relation ($M$ in \MS, see text).}
  \label{luminm}
\end{figure}

Figure \ref{ledd} shows luminosity $L_0$ in terms of $L_{\mathrm{Ed}}$
as a function of $\mu^2M$ [solid curve specified by Eq.$\,$(\ref{Lbetas})].
The curve virtually holds for all the
three energy generation modes --- a small difference for the pp-chain 
at $\mu^2M <10$ is indistinguishable on the scale of the figure.
For given $\mu^2M$, the Eddington model is always over-luminous 
(by a factor of two at $\mu^2M <10$). 

For the sake of practical use, it is worth to rewrite the mass-luminosity 
relation in terms of variable $\lambda$:
\begin{equation}
 \lambda\,\equiv\,\mu^2\frac{M}{\MS}\,\frac{L_0}{L_{\mathrm{Ed}}}\, =\,
 \frac{\mu^2\kappa_0 L_0}{4\pi cG\MS}=\xmn{1.5426}{-5}\mu^2(1+\mbox{X})
  \frac{L_0}{\LS}\, ,\label{lambd}
\end{equation}
Connecting the asymptotics for small and large $\mu^2M$-values by
a cubic spline that ensures the continuity of the function and its
first derivatives (open circles in Fig.~\ref{luminm}) one obtains
the following analytical approximation:  
\begin{eqnarray}
  \lg(\mu^2M) \aceqva
   0.9347\, +\, \frac{1}{3}\lg\lambda\, ,\qquad\qquad\qquad\qquad\qquad\quad
   (\lg\lambda\leqslant -1.7) ,\nonumber\\
  \lg(\mu^2M) \aceqva
   1.015209+0.449843\lg\lambda +0.0534969\lg^2\lambda\label{masslap}\\
     & & +\, 0.00754094\lg^3\lambda\, , \!
  \qquad\qquad\qquad\quad\;\;(-1.7 <\lg\lambda < 1.7) ,\nonumber\\
  \mu^2M \aceqva 10.125\left(1 + 0.0987654\lambda +\sqrt{1+0.1975\lambda}\right),
  \;\;\; (\lg\lambda\geqslant 1.7).\nonumber
 \end{eqnarray}
 \begin{table}
  \centering
  \caption{Integral properties of selected models}\label{integr}
  \vspace*{2mm}
 \begin{tabular}{l|llllllll}\hline\hline
 $\mu^2M$\!\!                & \multicolumn{1}{c}{0}   & \multicolumn{1}{c}{10}     
 &\multicolumn{1}{c}{30} & \multicolumn{1}{c}{100} & \multicolumn{1}{c}{300} 
 &\multicolumn{1}{c}{1000} & \multicolumn{1}{c}{4000}
 &\multicolumn{1}{c}{Edd}\\ \hline 
 $E_{\mathrm{g}}\, ^\ast$\!\! & -1.227 & -1.206 & -1.205 & -1.254 & -1.318 & -1.383 
 & -1.435 & -1.5\\
 $E_{\mathrm{T}}$\!\!    & 0.6137 & 0.6969 & 0.8350 & 1.016 & 1.166 & 1.292 & 1.387 
 & 1.450\\
 $I\, ^\dagger$\!\!      & 0.1561 & 0.1616 & 0.1630 & 0.1536 & 0.1412 & 0.1302 & 0.1222
 & 0.113\\
 $I_{\mathrm{conv}}$\!\! & 0.0139 & 0.0439 & 0.0908 & 0.1247 & 0.1309 & 0.1271 & 0.1214
 & \multicolumn{1}{c}{---}\\
 $t_{\mathrm{s}}$\!\!    & 2.438  & 2.517  & 2.580  & 2.624  & 2.649  & 2.665  & 2.677
 & 2.723\\
 $t_{\mathrm{conv}}$\!\! & 0.269  & 0.433  & 0.619  & 0.843  & 1.044  & 1.251  & 1.472
 & \multicolumn{1}{c}{---}\\
 $\tau_{\mathrm{c}}$\!\! & 1.441  & 1.426  & 1.475  & 1.664  & 1.896  & 2.132  & 2.329
 & 2.587\\
 $\langle\gamma\rangle$\!\!  & 5/3   & 1.532 & 1.453 & 1.401 & 1.3733 & 1.3555 & 1.3445
 & 1.345\\
 $\omega$\!\!                & 2.804 & 2.109 & 1.629 & 1.288 & 1.058  & 0.840 & 0.627
 & 0.667\\
 \hline\hline
\multicolumn{8}{l}{$^\ast$ \footnotesize $E_{\mathrm{g}}$ attains a maximum 
$-1.2005$ at $\mu^2M=18.5$}\\
\multicolumn{8}{l}{$^\dagger$ \footnotesize $I$ attains a maximum $0.1636$ at 
$\mu^2M=21$}
\end{tabular}
 \end{table}
 Equation~(\ref{masslap}) allows to estimate $M$ when the luminosity 
 $L_0$ and the representatives of composition ($\mu$ and $X$) are known. 
 On the contrary, in order to estimate $L_0$ for given $M$ and composition 
 one can either solve Eq.~(\ref{masslap}) for $\lambda$ or use 
 the following accurate approximation:
\begin{eqnarray}
  \lambda \aceqva
   0.00157\left(\mu^2M\right)^3,\qquad\qquad\qquad\qquad\qquad\quad\quad
   \left(\mu^2M\leqslant 2.4\right),\nonumber\\
  \lg\lambda \aceqva
 -2.907029+3.552793\lg\left(\mu^2M\right)-0.7717945\lg^2\left(\mu^2M\right)
  \label{lapmass}\\
  & & +\, 0.078623\lg^3\left(\mu^2M\right),\!
  \qquad\qquad\qquad\quad\left(2.4<\mu^2M < 100\right),\nonumber\\
  \lambda\aceqva \mu^2M\left(1 -\frac{4.5}{\sqrt{\mu^2M}}\right),\;\;\;
  \qquad\qquad\qquad\qquad\quad\left(\mu^2M\geqslant 100\right).\nonumber
 \end{eqnarray}
 \begin{figure}[th]
  \centerline{\includegraphics[clip]{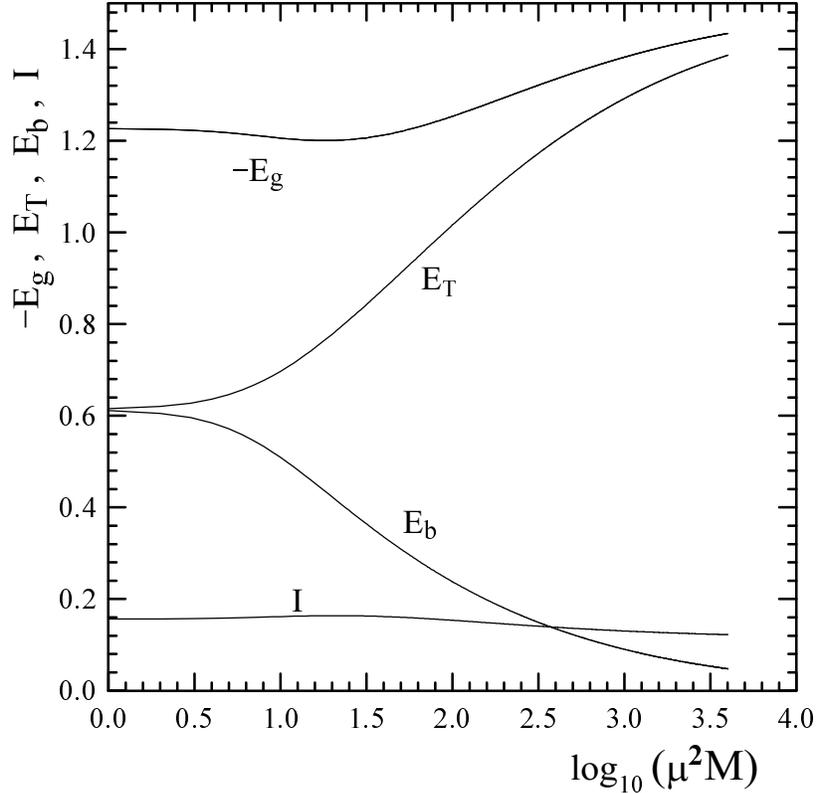}}
  \caption{Dimensionless gravitational $E_{\mathrm{g}}$ and thermal
  $E_{\mathrm{T}}$ energies, central moment of inertia $I$,
  and gravitational binding energy $E_{\mathrm{b}}$ versus $\mu^2M$.}
  \label{egitot}
  \end{figure}

Concluding this section we present in Table~\ref{integr}
a number of integral properties of stellar models such
as gravitational $E_{\mathrm{g}}$ and thermal $E_T$ energies 
(in units $GM^2/R$),
central moment of inertia of the whole star $I=\int^M_0\! r^2\,\mathrm{d}m$ 
and that of the convective
core $I_{\mathrm{conv}}$ (in units $MR^2$), the time sound takes to 
propagate from the center up to the surface 
$t_{\mathrm{s}}=\int^R_0\!\frac{\mathrm{d}r}{\sqrt{\gamma P/\rho}}$
and to the convective core boundary $t_{\mathrm{conv}}$
(in units $\sqrt{R^3/(GM)}\,$),
column density of the star $\tau_{\mathrm{c}}=\int^R_0\!\rho\,\mathrm{d}r$ 
(in units $M/R^2$), average adiabatic index 
$\langle\gamma\rangle=\frac{\int^M_0\!\gamma\, (P/\rho)\,\mathrm{d}m}
{\int^M_0\!(P/\rho)\,\mathrm{d}m}$, and an estimate of fundamental angular 
frequency 
$\omega =\sqrt{(3\langle\gamma\rangle -4)|E_{\mathrm{g}}|/I}$ 
of radial pulsations (in units $\sqrt{GM/R^3}\,$).

The last column in Table~\ref{integr} presents the properties of 
the Eddington standard model at $\mu^2M =4000$. Note that for the standard
model the dimensionless $E_{\mathrm{g}}$, $I$ and $\tau_{\mathrm{c}}$ 
are determined by the polytrope $n=3$ structure and do not
depend on $\mu^2M$. 

Figure~\ref{egitot} shows $E_{\mathrm{g}}$, 
$E_{\mathrm{T}}$, $I$, and gravitational binding energy
$E_{\mathrm{b}}=-E_{\mathrm{tot}}=-(E_{\mathrm{g}}+E_{\mathrm{T}})$
as functions of $\mu^2M$.

 \section{Comparison with detailed models}
\begin{figure}[ht]
    \centerline{\includegraphics[clip]{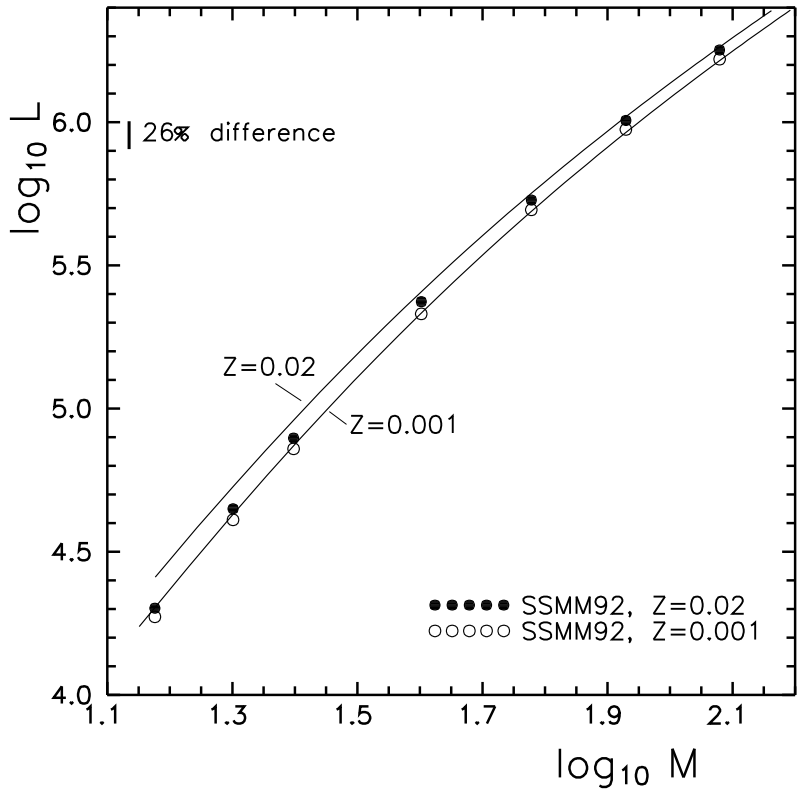}}
  \caption{The mass-luminosity relation (solid lines) in comparison 
           with the detailed models of Schaller et al. (1992) 
           (filled and open circles). 
           $M$ and $L$ are in solar units.}
  \label{maedlmo}
\end{figure}

\begin{figure}[ht]
     \centerline{\includegraphics[clip]{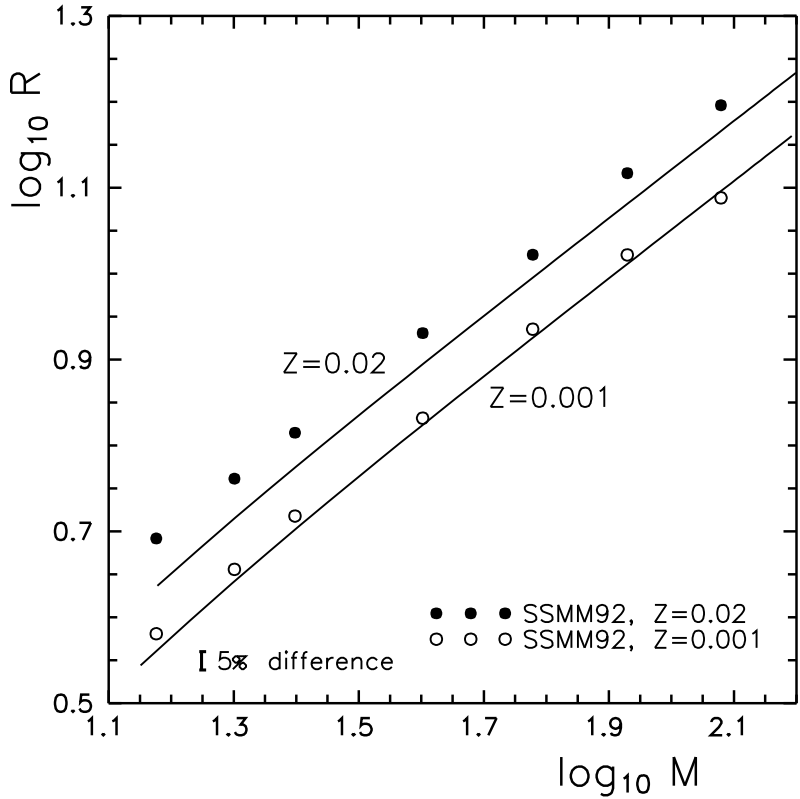}}
  \caption{The mass-radius relation (solid lines) in comparison with
           the detailed models of Schaller et al. (1992) (filled and 
           open circles). $M$ and $R$ are in solar units.}
  \label{maedrmo}
\end{figure}

\begin{figure}[hb]
  \centerline{\includegraphics[clip]{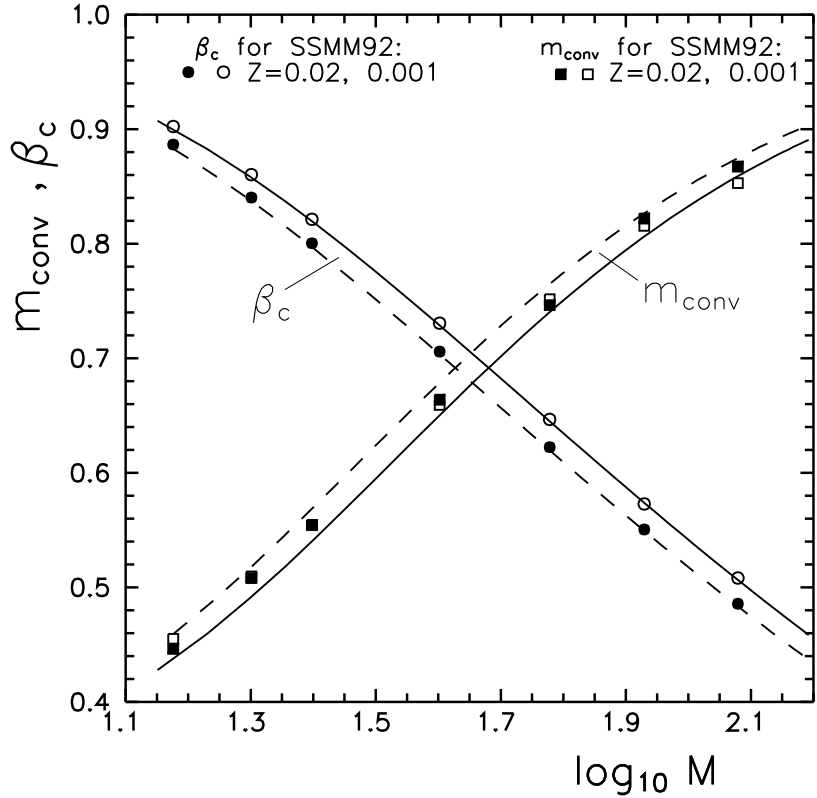}}
  \caption{Dimensionless mass of the convective core $m_{\mathrm{conv}}$
           and central value of $\beta$ (solid and dashed lines)
           in comparison with Schaller et al. (1992) (filled and open circles
           and squares). Solid and dashed lines are for $Z=0.001$ and $Z=0.02$,
           respectively; $M$ is in \MS.}
  \label{maedbcq}
\end{figure}

\begin{figure}[ht]
     \centerline{\includegraphics[clip]{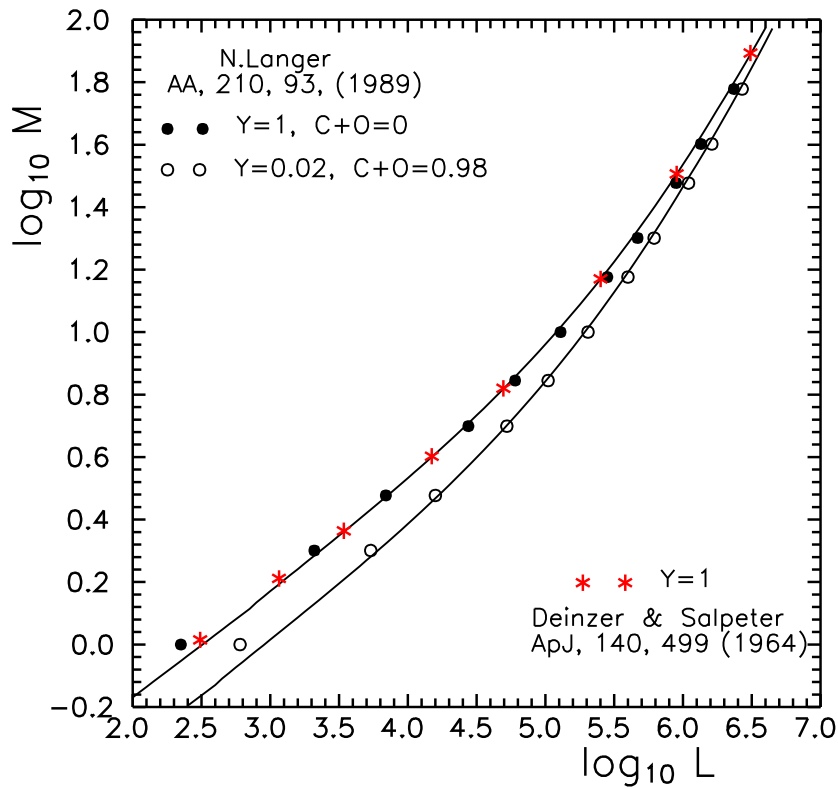}}
  \caption{The mass-luminosity relation for the helium and carbon-oxygen
           Wolf-Rayet stars (solid lines) in comparison with
           the detailed models of Langer (1989) (filled and open circles)
           and Deinzer and Salpeter (1964) (asterisks).
           $M$ and $L$ are in solar units.}
  \label{wrlangr}
\end{figure}

 In order to demonstrate potentialities of the similarity theory we compare
 our results with detailed models of massive main sequence stars calculated
 by  Schaller et al. (1992) and the models of helium and carbon-oxygen 
 stars (Wolf-Rayet stars) studied by Langer (1989) and 
 Deinzer and Salpeter (1964). Figures \ref{maedlmo}--\ref{wrlangr} display
 such a comparison.
Solid curves in Fig.~\ref{maedlmo} are obtained with the aid of 
Eq.~(\ref{lapmass}) for two specified by Schaller et al. (1992)
compositions: X=0.680, Y=0.300, Z=0.020 (upper curve) and
X=0.756, Y=0.243, Z=0.001 (lower curve).
One can observe an excellent agreement for low  metallicity 
$(Z=0.001)$ and a fairly good coincidence for $Z=0.02$. 
In the latter case our models are slightly over-luminous 
(by 25\% at $M=\NMS{15}$).  This natural result 
can be interpreted as on average a 25\% contribution to 
the opacity coming from sources other than the electron scattering.

 The mass-radius relation for the same models is shown in Fig.~\ref{maedrmo}.
 The detailed models have systematically larger radii 
 (typically by $\sim 5\%$ for Z=0.02) as compared with our models (solid lines).
 This can be explained by a noticeable increase in opacity in stellar envelope 
 owing to the absorption in atomic spectral lines (Imshennik, Nadyozhin, 1967).
 The solid lines were calculated with the aid of Eqs.~(\ref{Rad}) and (\ref{LC1m})
 assuming the CNO-cycle as the main energy source with the energy generation
 law taken from Caughlan, Fowler (1988) and approximated by
  \begin{equation}
  \varepsilon_{\mathrm{CNO}}\, =\,\xmn{8.43}{-3}X_{\mathrm{CNO}}X\rho\, 
  T^{16}_7\;\erggs ,\quad (2.2\leqslant T_7\leqslant 3.6)\, ,\label{ECNO}
  \end{equation}
  where $T_7\equiv T/10^7\mathrm{K}$ and $\rho$ is in \gccm.
  Within the indicated limits for $T_7$, the accuracy of  
  Eq. (\ref{ECNO}) is better than 10\%. 

  Similar approximation for 
  the $3\alpha$-reaction is given by the following power law
  \begin{equation}
  \varepsilon_{3\alpha}\, =\,\xmn{4.95}{-38}\mathrm{Y}^3\rho^2\, T^{30}_7\;
  \erggs ,\quad (T_7 >8)\, .\label{ALFA}
  \end{equation}
   The accuracy of the approximation is better than 10\% at $T_7 >8$. 
   
   The structural properties of our models are also in a good agreement
   with the detailed models as it is exemplified in Fig.~\ref{maedbcq}
   for central value of $\beta$ and the convective core mass fraction
   $m_{\mathrm{conv}}$. 

Figure~\ref{wrlangr} shows the mass-luminosity relation for pure helium
stars (Y=1) calculated by  Deinzer and Salpeter (1964) and Langer (1989),
and for carbon-oxygen Wolf-Rayet stars, with the mass fractions of
carbon, oxygen, and helium
X$_{\mathrm{C}}$=0.113, X$_{\mathrm{O}}$=0.867, Y=0.02, respectively 
(Langer, 1989),
in comparison with the mass-luminosity relation for our models determined
by Eq.~(\ref{masslap}) (solid lines). One has to specify the hydrogen 
mass fraction X=0 in Eq. (\ref{lambd}) for $\lambda$ and make use of
the following expression for the mean molecular mass $\mu$:
\begin{equation}
 \mu\, =\,\frac{48}{36\mathrm{Y}+28\mathrm{X}_\mathrm{C}+27\mathrm{X}_\mathrm{O}}
 \, ,\qquad (\mathrm{Y}+\mathrm{X}_\mathrm{C}+\mathrm{X}_\mathrm{O}=1)
 \, .\label{muCO}
 \end{equation}

Our results are in a very good agreement with the detailed models even 
for such a small mass as \NMS{1}. 

 \section{Discussion and conclusions}
 It is interesting to apply the mass-luminosity relation given by
 Eqs.~(\ref{masslap}) and (\ref{lapmass}) to the very massive stars observed
 in the Milky Way, in Magellanic Clouds and in a number of nearby 
 resolved galaxies. There exist comprehensive observational data for 
 several dozens of such stars (see, for instance, Figer et al., 1998; 
 Puls et al., 1996; Humphreys, Davidson, 1994; Sandage, Tammann, 1974).
 A detailed analysis of the efficiency of the similarity theory in
 deriving the properties of such stars from observations is in need of 
 a special paper. Here as an example we consider the Pistol star 
 studied in detail by Figer et al. (1998). 
 Assuming that the star was initially of solar 
 composition (X$=$0.707, Y$=$0.274, Z$=$0.019; Anders, Grevesse, 1989)
 and had the luminosity $L_0=10^{6.7\pm 0.5}\LS$ (Najarro, Figer, 1999),
 one can estimate from  Eq.~(\ref{masslap}) its initial mass to be
 $M=116$, $246$, and \NMS{595} for $L_0=10^{6.2}$, $10^{6.7}$, 
 and \NLS{10^{7.2}}, respectively. The corresponding initial radii
 derived from Eqs. (\ref{Rad}) and (\ref{LC1m}) turn out to be
 $R=14$, $22$, and \NRS{36} for the CNO energy generation rate 
 given by Eq. (\ref{ECNO}). The range of masses \NMS{(116-595)} 
 corresponds to the range $(44-230)$ in the parameter $\mu^2M$
 since $\mu=0.618$ for the solar composition.
 Such stars have luminosity of $(0.4-0.7)L_{\mathrm{Ed}}$ 
 (Fig. \ref{ledd}) and very large convective cores: 
 $m_{\mathrm{conv}}=0.86-0.976$ [Eq.~(\ref{mabov}) or Fig.~\ref{convec}].
 Estimating the Pistol star initial mass we have implied that  
 its luminosity has not changed appreciably during the evolution.
 The star seems to be in a state close 
 to the exhaustion of hydrogen in the convective core --- i.e., 
 it is about to leave the Main Sequence. According to Schaller et al.
 (1992), an increment in the luminosity $\Delta L_0$ within 
 the Main Sequence strip decreases with increasing mass $M$. 
 Since for $M=\NMS{120}$ (solar metallicity) $\Delta L_0\approx 25\%$,
 initially the Pistol star might have been by $\sim 10\%$ 
 less massive than the estimates stated above.
 
 The properties of stellar structure obtained here are in a fair 
 agreement also with the detailed models of the very massive,
 \NMS{(100-250)}, initially zero-metallicity Pop III stars calculated
 in the UCSC astrophysical group (Woosley, 2005), especially when
 the stars are definitely settled at their Main Sequence in the
 state of thermal equilibrium. 

 The similarity theory of the structure of very massive stars
 formulated here allows to obtain the following important results:
 \begin{enumerate}
 \item  A simple approximation formula for the mass-luminosity
  relation, given by Eqs.~(\ref{lambd}),$\,$(\ref{masslap}),
  and (\ref{lapmass}),
  is valid for different chemical compositions and virtually 
  does not depend on the energy generation mode (pp-chain, CNO-cycle,
  or $3\alpha$-reaction). 
 \item The overall structure of very massive stars, described in dimensionless
  variables listed in Eq. (\ref{dimlessvar}), depends only on the parameter
  $\mu^2M$ where $\mu$ is the mean molecular mass. Practically,
  this structure turns out to be independent of the energy generation
  mode --- be it the CNO-cycle, $3\alpha$-reaction, or pp-chain. 
  \item  Although with increasing $\mu^2M$ the stellar structure approaches
  that of the Eddington standard model, the convergence proves to be  
  rather slow. Even for $\mu^2M=4000$ there are still noticeable discrepancies
  in some stellar parameters. For instance, the central dimensionless pressure 
  and density are respectively by about 20\% and 15\% lower than for 
  the Eddington model (Fig.~\ref{rhopcen}). 
 \end{enumerate}
 
 \subsection*{Acknowledgements}
 One of us (DKN) has a pleasure to thank the Max-Planck-Institut 
 f\"ur Astrophysik for financial support and hospitality.
 The work was partly supported by the Russian Foundation for
 Basic Research (project no. 04-02-16793-a).
 
\subsection*{References}

\begin{itemize}
\item[] Anders E., Grevesse N.\\
       Geochim. Cosmochim. Acta 1989, {\bf 53}, 197.

\item[] Biermann L. Zs. f\"ur Astrophysik 1931, {\bf 3}, 116.

\item[] Caughlan, G.R., Fowler, W.A.\\
        Atomic Data Nucl. Data Tables 1988, {\bf 40}, 283.

\item[] Chandrasekhar S. 
        {\em An Introduction to the Study of Stellar Structure\/}, 
        Chicago: Univ. of Chicago Press, 1939.

\item[]  Chiu H.--Y.
         {\em Stellar Physics\/}, vol. 1, Blaisdell Pub.,\\
         Massachusetts-Toronto-London, 1968.

\item[]  Cowling T.G.
         Mon. Not. RAS 1934, {\bf 94}, 768.

\item[]  Cox J.P., Guili R.T.
         {\em Principles of stellar structure\/}, vol. 2, \\
         Gordon and Breach, New York, 1968.

\item[]  Deinzer W., Salpeter E.E.
         Astrophys. J. 1964, {\bf 140}, 499.
         
\item[]  Dibai E.A., Kaplan S.A.,
         {\em Razmernosti i podobie astrofizicheskikh ve\-li\-chin\/} 
         (Dimensions and similarity of astrophysical quantities),\\
          Nauka Pub., Moscow, 1976 (in Russian).
         
\item[]  Eddington A.S.
         {\em The Internal Constitution of the Stars\/},\\
         Cambridge, 1926.
         
\item[]  Figer D.F., Najarro F., Morris M., McLean I.S., Geralle T.R.,\\
         Ghez A.M., Langer N.
         Astrophys. J. 1998, {\bf 506}, 384.
                 
\item[]  Henrich L.R.
         Astrophys. J. 1943, {\bf 98}, 192.

\item[]  Humphreys R.M., Davidson K.\\
         Publ. Astron. Soc. Pacific 1994, {\bf 106}, 1025.

\item[]  Imshennik V.S., Nadyozhin D.K.
         Astron. Zh. 1967, {\bf 44}, 377.\\
         (Sov. Astr. 1967, {\bf 11}, 297).

\item[]  Imshennik V.S., Nadyozhin D.K.
         Astron. Zh. 1968, {\bf 45}, 81.\\
         (Sov. Astr. 1968, {\bf 12}, 63).

\item[]  Kippenhahn R., Weigert A.
         {\em Stellar structure and evolution\/},\\
         Springer-Verlag, Berlin-Heidelberg, 1990.

\item[]  Langer N. 
         Astron. Astrophys. 1989, {\bf 210}, 93.
 
\item[]  Najarro F., Figer D.F. \\
         Astrophys. Space Sci. 1999, {\bf 263}, 251.
                     
\item[]  Naur P., Osterbrock D.E.
         Astrophys. J. 1953, {\bf 117}, 306.

\item[]  Puls J., Kudritzki R.-P., Herrero A., et al.\\
         Astron. Astrophys. 1996, {\bf 305}, 171.
         
\item[]  Sandage A., Tammann G.A. 
         Astrophys. J. 19474, {\bf 191}, 603.
         
\item[]  Schaller G., Schaerer D., Meynet G., Maeder A.\\
         Astron. Astrophys. Suppl. 1992, {\bf 96}, 269.
        
\item[]  Schwarzschild M.
         {\em Structure and Evolution of Stars\/},\\
         Princeton Univ. Press, Princeton, New Jersey, 1958.

\item[]  Sedov L.I. {\em Similarity and dimensional methods 
         in mechanics\/} \\
         Academic Press, New York, 1959
         (4th Russian ed.). See also \\
         the 8th revised edition, Nauka Pub., Moscow, 1977 (in Russian).

\item[]  Str\"omgren B. Handbuch der Astrophysik 1936,
         {\bf 7}, 121.
 
\item[]  Woosley S.E. 2005, private communication.
        
\end{itemize}

\end{document}